\input{aipcheck}

\documentclass[
    ,final            
  ]
  {aipproc}

\layoutstyle{6x9}

\begin{document}

\title{Feeding and Feedback in nearby AGN  from Integral Field Spectroscopy}

\classification{98.62.Js, 98.62.Mw, 98.62.Nx}
\keywords      {Galaxies: nuclei; Galaxies:active; Galaxies:ISM}

\author{Thaisa Storchi-Bergmann}{address={Instituto de F\'isica, Campus do Vale, CP\,15051, Universidade Federal do Rio Grande do Sul, Porto Alegre, RS, Brasil}}

\begin{abstract}
I report results of recent integral field spectroscopy of the inner few hundred parsecs (pc) around nearby Active Galactic Nuclei (AGN) at a sampling of a few pc, obtained with the Gemini Telescopes. In the lowest activity AGNs, it is possible to observe inflows in ionized gas along nuclear spirals and filaments. In more luminous AGN inflows have been observed also in hot molecular gas (H$_2$) emission in the near-IR. In most cases the H$_2$ kinematics is dominated by circular rotation in the plane around the nucleus, tracing the AGN feeding. The ionized gas, on the other hand, traces the AGN feedback. Its kinematics shows two components: (1) one originating in the plane, and dominated by circular rotation; (2) another outflowing along the Narrow-Line Region (NLR) whose flux distribution and kinematics frequently correlate with structures seen in radio maps. 
Mass outflow rates along the NLR range from 10$^{-2}$ to 1\,M$_\odot$\,yr$^{-1}$, corresponding to 10--100 times the accretion rate to the AGN, indicating that most of the NLR gas mass has been entrained from the galaxy plane. The average kinetic power of the NLR outflows is $\approx$\,10$^{-4}$ times the bolometric luminosity.

\end{abstract}

\maketitle


\section{Introduction}

Outflows are ubiquitous among AGN \citep{veilleux05}, and are observed from radio \citep{morganti05}, through 
X-rays \citep{chelouche05}. These outflows provide the feedback necessary to regulate the growth of galaxies and the M vs. $\sigma$ relation \citep{matteo05}. The origin of the outflows seem to be radio ejecta from the funnel of the accretion disks and/or accretion disk winds \citep{elvis00}.
But in order to trigger the nuclear activity and produce the resulting outflows it is first necessary to feed the nuclear supermassive black hole (hereafter SMBH). Inflows are, nevertheless, seldom observed. Where should we look for them? In \citep{sl07} we have shown that in HST optical images of early-type AGNs and a matched sample on non-AGN,  nuclear filaments or spirals are present in all AGN but in less than a third of the non-AGN, suggesting that such strucutures map channels which feed the SMBH in active nuclei.

\section{Feeding}

In order to look for kinematic signatures of inflows along the nuclear spirals, my group began a project to map the gas kinematics in the inner few hundred pc of nearby AGN. We have used the Integral Field Spectrographs available at the Gemini Telescopes, namely the Integral Field Unit of the Gemini Multi-Object Spectrograph (GMOS IFU), in the optical, and the Near-Infrared Integral Field Specgtrograph (NIFS) in the near-infrared.

{\it\bf Optical Observations.} In \citet{fathi06} we have used the GMOS IFU to observe the inner 7$^{\prime\prime}\times\,15^{\prime\prime}$ of NGC\,1097, along the galaxy major axis, in the spectral region $\sim\,6000-7000$\AA. We have obtained the gaseous kinematics in the emission lines of H$\alpha$+[NII]$\lambda6548,84$\AA. After sutracting the velocity field of a circular disk model from the data, we found non-circular motions in association with a nuclear spiral, revealing streaming motions towards the center with velocities of $\sim$\,50\,km\,s$^{-1}$. In \citet{sb07} we showed a similar result for the galaxy NGC\,6951, and calculated a mass inflow rate in ionized gas of $\approx\,10^{-3}$M$_\odot$\,yr$^{-1}$. Coincidentally, this is about the mass accretion rate necessary to feed the SMBH in these low-luminosity AGN. Most probably, the total mass inflow is dominated by non-emitting neutral and molecular gas and we are just seeing the ionized ``skin'' of the actual flow.

We have tried to look for inflows in higher luminosity AGN, but in these cases the kinematics is dominated by outflows (see below). We have recently found circumnuclear gas inflow in M\,81, another low-luminosity AGN, for which we were able to measure the stellar kinematics and have subtracted it from the gas kinematics. The results were residuals in the form of blueshifts of up to $\sim\,-100$\,km\,s$^{-1}$ in the far side of the galaxy and redshifts in the near side, revealing streaming motions of the gas towards the center along the minor axis of the galaxy \citep{schnorr10}.

\begin{figure}
\includegraphics[height=.4\textheight]{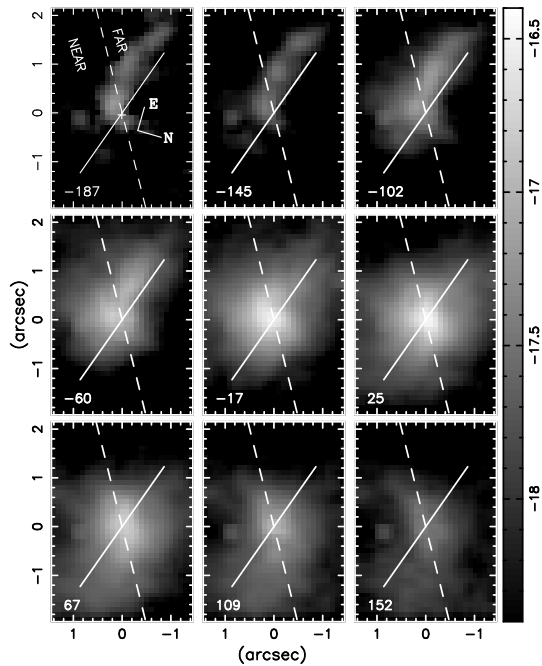}
\caption{Feeding: channel maps along the  H$_2\lambda$2.1213\,$\mu$m emission-line profile of the nuclear region of the Seyfert galaxy NGC\,4051, revealing inflows towards the nucleus. The numbers at bottom left of each panel show the central velocities after subtraction of the systemic velocity. From \citep{riffel08}.}
\end{figure}

{\it\bf Near-infrared Observations.} We have found inflows in the near-infrared in NIFS observations of the inner $\sim$\,300\,pc of the active galaxy NGC\,4051 \citep{riffel08}. These inflows were observed in channel maps along the H$_2\lambda$2.1213\,$\mu$m emission-line profile. Blueshifts are observed in the far side of the galaxy, while redshifts are observed in the near side, as shown in Fig. 1. If the gas is in the plane of the galaxy, this kinematics means inflows towards the center. We have calculated the mass inflow rate in the molecular gas, and found low values, which corresponds to only $\approx\,1$\%  of the mass accretion rate. Much more gas is necessary in order to feed the SMBH what leads us to the conclusion that we are observing again only the ``hot skin'' of the molecular gas flow. Estimated values for the ratio of cold to hot molecular gas mass in the nuclear region of AGN host galaxies range from 10$^5$ to 10$^6$ \citep{dale05}. In NIFS observations of NGC\,4151 \citep{sb10} and in Mrk\,1066 \citep{riffel10}  we found molecular gas rotating around the nucleus in compact disks, which may be fuelling the accretion disk inside. 


\section{Feedback}

{\it\bf Optical Observations.}
We have used GMOS-IFU observations to map both the stellar and gas kinematics of 6 nearby Seyfert galaxies \citep{barbosa09}. We have used the Calcium Triplet at $\sim\,$8500\AA\ to map the stellar kinematics and the [SIII]$\lambda$9069\AA\ emission line to map the gas kinematics. The stellar kinematics is dominated by circular rotation, while the gas kinematics -- as measured from the centroid of the emission-line profiles -- shows both circular rotation and outflows of $\sim$\,200\,km\,s$^{-1}$.  We found that the gas emission is not restricted to outflowing region, and the gas which is not outflowing shows circular rotation similar to that of the stars, which favors an origin for this gas in the plane of the galaxy. It can thus be concluded that, the AGN ionizing radiation escapes to regions beyond the narrow-line region (NLR) and in the equatorial plane of the outflow, indicating that the obscuring torus  is probably clumpy, as suggested by recent models \citep{elitzur06}.

The outflows are in most cases associated with  jets and blobs observed in radio images, when available, suggesting that at least in our sample the NLR outflow is produced by radio jets pushing the circumnuclear ambient gas. This is supported also by the values estimated for the mass outflow rate, in the range 10$^{-3}-10^{-2}$ M$_\odot$\,yr$^{-1}$. These values are  $\sim$\,10 times the accretion rate to feed the SMBH, indicating that the origin of the NLR gas is not the AGN itselft but is ambient gas entrained by an AGN outflow. The kinetic power of the outflow is $\sim\,10^{-4}$ times the bolometric luminosity. 

Channel maps along the [SIII] emission-line profile show that the highest velocity gas is observed at the nucleus, contrary to results of a few previous studies for other galaxies which suggest that there is acceleration along the NLR.

{\it\bf Near-infrared Observations.}
We have mapped outflows in the near-infrared using NIFS \citep{sb10,riffel10} and the IFU of the Gemini Near-Infrared Spectrograph \citep{riffel06}. The main results of these studies are similar to those obtained in the optical: (1) ionized gas emission is not restricted to the NLR; (2) the gas in the NLR is outflowing, with centroid velocities of a few hundred km\,s$^{-1}$, and is usually correlated with structures observed in radio maps; (3) the gas which is not outflowing is in the plane of the galaxy and is usually observed all around the nucleus, indicating escape of ionizing radiation from the AGN even at the base of the outflow; (4) channel maps usually show high velocities at the nucleus, and do not support acceleration along the NLR. This can be observed, for example, in channel maps of the nuclear region of NGC\,4151, shown in the paper by Sim\~oes Lopes et al. in these Proceedings.

Fig. 2  shows channel maps in the [FeII]$\lambda$1.64$\mu$m emission line  for the galaxy Mrk\,1066, revealing outflows which correlate with the radio emission (green contours).


\begin{figure}
\includegraphics[height=.4\textheight]{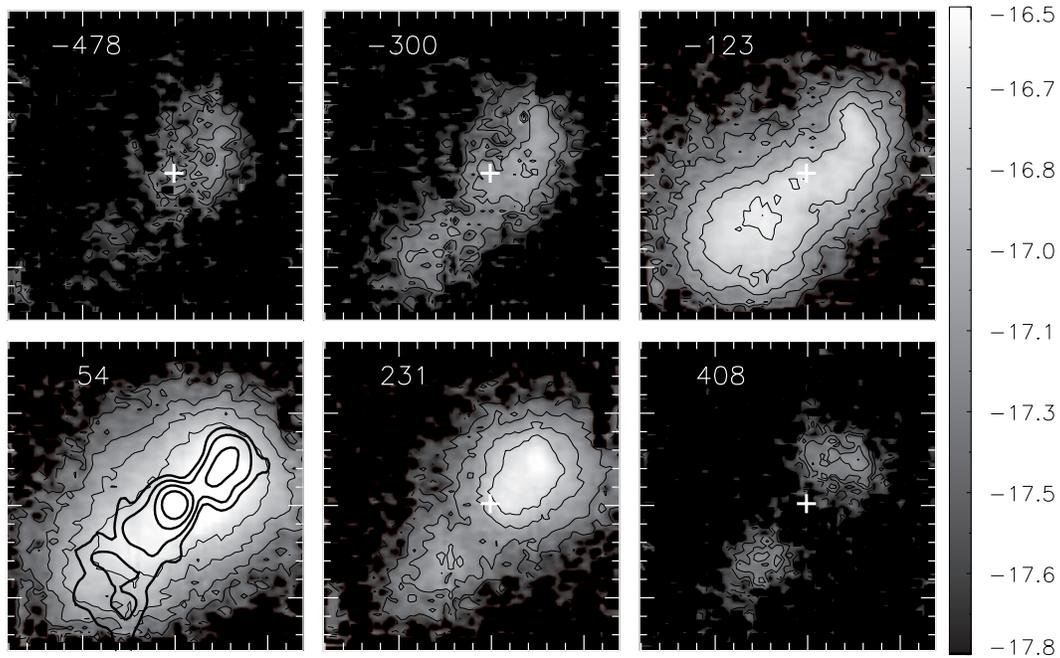}
\caption{Feedback: channel maps along the [FeII]$\lambda$1.64$\mu$m emission-line profile of the nuclear region of the Seyfert galaxy Mrk\,1066, from \citep{riffel10}. Heavy contours are from a radio image \citep{nagar99}. The numbers at the top left of each panel show the central velocities after subtraction of the systemic velocity. North is up and East to the left, and the large tick marks are separated by 1 arcsec (224\,pc at the galaxy).}
\end{figure}

\end{document}